\documentclass[a4paper,UKenglish]{lipics-v2021}
\usepackage{graphicx}
\usepackage{listings}
\usepackage{booktabs}
\usepackage{makecell}

\title{From Business Requirements to Test Assertions: Evaluating LLM-Generated Oracles on Real Bugs}

\titlerunning{LLM-Generated Test Oracles from Business Requirements}

\author{Tiancheng Ma}{Department of EECS, University of Tennessee, Knoxville, TN, USA}{tma10@vols.utk.edu}{}{}

\author{Nasir U. Eisty}{Department of EECS, University of Tennessee, Knoxville, TN, USA}{neisty@utk.edu}{}{}

\authorrunning{T. Ma and N. U. Eisty}

\Copyright{Tiancheng Ma and Nasir U. Eisty}



\relatedversion{}

\EventEditors{To be completed by editors}
\EventNoEds{2}
\EventLongTitle{20th ACM/IEEE International Symposium on Empirical Software Engineering and Measurement (ESEIW 2026)}
\EventShortTitle{ESEIW 2026}
\EventAcronym{ESEIW}
\EventYear{2026}
\EventDate{To be announced}
\EventLocation{To be announced}
\EventLogo{}
\SeriesVolume{0}
\ArticleNo{0}
\DOIPrefix{}

\nolinenumbers

\begin{document}
\makeatletter\def\@oddfoot{}\def\copyrightline{}\makeatother

\maketitle

\begin{abstract}
\textbf{Background.} The oracle problem (determining the correct expected outcome for a test) remains a major bottleneck in automated testing, and is increasingly relevant as non-experts rely on AI-generated code they cannot reliably validate.
\textbf{Objective.} We study whether large language models (LLMs) can generate \emph{generalizable} test oracles directly from natural-language business requirements, without access to source code or example input--output pairs.
\textbf{Method.} We propose a reproducible, requirement-driven pipeline grounded in Defects4J. For each of 10 real bugs from Defects4J Lang (Bugs 1 and 3--11), we (i) extract behavioral changes via buggy/fixed diffs, (ii) manually translate the change into a business requirement, (iii) construct a requirement-derived oracle (REQ) as a gold standard, and (iv) prompt five LLMs (DeepSeek-V3, Gemma-3n, Llama-3, Mistral-7B, and Qwen-3) to generate Java oracle code. We evaluate oracle correctness and generalization under two targets: agreement with REQ and agreement with the system under test (SUT), reporting macro-averaged accuracy, precision, recall, and F1.
\textbf{Results.} LLMs achieve non-trivial generalization but with substantial bug- and model-level variance. Generated oracles align more closely with REQ than with SUT, and correlations between requirement technicality/ambiguity ratings and oracle accuracy are weak with wide confidence intervals.
\textbf{Conclusion.} No detectable linear relationship exists between requirement properties and oracle accuracy in this dataset, suggesting that pretraining coverage and the semantic specificity of the required behavior dominate oracle correctness. As a pilot proof of concept, these findings are preliminary and are intended to establish feasibility and motivate larger-scale empirical investigation.
\end{abstract}

\section{Introduction}
\textbf{Context.} A successful software project depends not only on correct and verified implementation, but on the accurate translation of real-world business requirements into reliable software behavior. In practice, business requirements are written in natural language, which is well known to be prone to ambiguity, incompleteness, and inconsistencies \cite{necula2024review}.
 These characteristics make it difficult for developers to precisely interpret the intended behavior and implement code that fully satisfies stakeholder expectations.

Moreover, delivering high-quality software requires not only correct implementation but also extensive testing, since no developer can guarantee that their code is free of defects. Effective testing, however, demands substantial effort: developers must create large numbers of test cases along with corresponding test oracles, the expected outputs used to determine correctness. Producing these oracles often requires deep domain knowledge, significant human labor, and in many scenarios, the correct expected output may be unclear or even inaccessible for complex test cases. As a result, oracle construction has long been recognized as one of the most costly and difficult components of software testing. This challenge is especially relevant when testers only have access to high-level business requirements, rather than source code, formal specifications or example input-output pairs.

At the same time, the landscape of software development is rapidly shifting with the rise of AI. Both technical developers and non-technical professionals increasingly rely on AI tools to generate code, analyze data, or prototype new software features. Yet most of these users cannot reliably verify whether the AI-generated code is correct or practical. They often resort to ad-hoc verification or subjective judgment, but such methods are slow, inconsistent, and heavily influenced by an individual’s background and technical expertise. Recent studies further show that non-technical business professionals often struggle to identify flaws in AI-generated analyses, even when the errors require no programming knowledge to detect \cite{virk2025nonprogrammersassessingaigeneratedcode}. This gap between AI-generated output and users’ ability to verify correctness highlights the growing need for automated, reliable mechanisms for validating system behavior.

Taken together, these observations reinforce the long-standing oracle problem~\cite{weyuker1982testing,barr2015oracle}: determining the correct expected output is frequently the bottleneck of automated testing. This motivates investigating whether modern LLMs can help automate oracle construction, enabling both experts and non-experts to verify AI-generated code more efficiently and effectively.

While prior research has explored using LLMs to generate test inputs from natural language requirements and infer oracles from existing pairs of inputs and outputs, very few studies have examined whether LLMs can generate reliable and generalized test oracles directly from business requirements~\cite{molina2025test}.

\textbf{Proposal \& Evaluation.} To investigate whether LLMs can generate reliable and generalizable test oracles from business requirements, we design a systematic and reproducible pipeline grounded in Defects4J. We adopt Defects4J as our primary dataset because it is actively maintained, widely used in software engineering research, and provides a curated collection of real, historical bugs with both buggy and fixed versions available. This allows us to construct controlled, realistic scenarios that reflect true software behaviors.

Our approach operates at the granularity of individual bugs from projects of Defects4J~\cite{just2014defects4j}. For each bug, we first check out the buggy and fixed versions of the project and compute the code differences between them, which provide the ground-truth behavioral change introduced by the fix. Using this information, we translate the observed change into a natural-language business requirement to intentionally mirror real-world development conditions, where requirements are expressed at a high level and rarely contain explicit code-level details.

Based on the business requirement, we then construct a standard test oracle that precisely captures the expected correct behavior after the fix. This oracle serves as the authoritative specification against which LLM-generated oracles can be evaluated. Next, we use the business requirement to construct prompts that are provided to five LLMs, asking them to generate their own test oracles without exposing any source code or diff information. This setup reflects a realistic use case in which LLMs operate purely from business-level descriptions, which would occur for many non-technical AI users.

To evaluate both human-constructed and LLM-generated oracles, we design a diverse set of test cases to measure not only correctness but also generalizability, whether the oracle captures the underlying intent of the requirement beyond the examples explicitly described.

\textbf{\textit{Contribution.}} We contribute \textbf{(i)} a reproducible, requirement-driven pipeline for evaluating LLM oracle generation from business requirements alone, grounded in Defects4J~\cite{just2014defects4j}, \textbf{(ii)} empirical results across five LLMs and ten real bugs with two oracle-correctness targets, and \textbf{(iii)} a correlation analysis of requirement attributes and oracle accuracy that informs future benchmark design. This work is scoped as a \emph{pilot proof of concept} limited to ten bugs from a single Defects4J project (Lang); results are preliminary and intended to establish feasibility, surface early empirical findings, and motivate future large-scale investigation across multiple projects and LLMs.

\textbf{Research Question.} 
To guide our study, we investigate the following research questions:
\begin{itemize}
    \item \textbf{RQ1:} How well can LLMs generate requirement-driven test oracles that generalize to unseen test cases?
    \item \textbf{RQ2:} How do LLM-generated oracles compare to the requirement-derived oracle (REQ) and the system-under-test (SUT) across different types of bugs?
    \item \textbf{RQ3:} How do the characteristics of business requirements, particularly technicality and ambiguity, influence the performance of LLM-generated oracles?
\end{itemize}

\section{Related Work}

Recent work on test oracle automation spans traditional specification-based methods, learning-based techniques, and emerging LLM-based approaches. Classical mechanisms such as assertions, contracts, and metamorphic relations remain limited in coverage and applicability, while newer formal and learning-driven methods improve behavioral completeness. With LLMs, researchers increasingly explore generating oracles from natural language and documentation, showing promising gains but still facing challenges in oracle quality, generalization, and reproducibility.

Pre-LLM oracle generation relied on formal specifications or learned behavioral models. Liu et al.'s Vibration Method \cite{liu2022automatic} automatically generates both test cases and oracles directly from pre- and post-conditions, introducing test generation criteria that ensure coverage of functional scenarios and their corresponding execution paths, an improvement over traditional partitioning-based techniques that often leave paths uncovered. The SEER framework \cite{ibrahimzada2022perfect} takes a fundamentally different approach by learning to distinguish passing from failing behavior through a joint embedding of unit tests and methods under test, eliminating the need for explicit assertions or formal specifications. Evaluated on more than 5,000 Java unit tests, SEER achieved 93\% accuracy, 86\% precision, and 90\% F1, with strong generalization to unseen projects.

With the rise of LLMs, researchers have explored generating oracles from documentation, code, and natural language. Molina et al. \cite{molina2025test} survey the landscape across 37 recent LLM-based studies, finding that most work targets assertion generation while fewer address invariants, metamorphic relations, or exception oracles; they identify oracle quality, reproducibility, and dataset bias as key open challenges. Khandaker et al.'s AugmenTest \cite{khandaker2025augmentest} generates oracles from documentation and developer comments rather than source code, comparing four prompting strategies on 142 Java classes; an extended prompting strategy achieved up to 30\% correctness compared to 8.2\% for TOGA, though RAG-based approaches underperformed, suggesting difficulties integrating structured retrieval with LLM inference. Hossain et al.'s TOGLL \cite{hossain2025togll} fine-tunes seven code LLMs with six prompt designs on 110 large-scale Java projects, producing 3.8$\times$ more assertion oracles and 4.9$\times$ more exception oracles than TOGA while detecting over 1,000 mutants missed by EvoSuite. 

On the test-input side, Hasan et al. \cite{hasan2025automatic} generate high-level test skeletons from natural-language scenarios, and Sch{\"a}fer et al. \cite{schafer2024empirical} demonstrate that LLM-generated tests achieve coverage competitive with EvoSuite but exhibit persistent difficulty with complex branching logic. Yuan et al. \cite{yuan2023chatgpt} evaluate ChatGPT for unit test generation and find that oracle correctness, not syntactic validity, is the primary bottleneck, directly motivating our focus on requirement-driven oracle generation.

Overall, while LLM-based test generation has advanced rapidly, generating correct, generalizable oracles directly from natural-language business requirements remains under-explored~\cite{molina2025test,barr2015oracle}. Existing work largely assumes access to source code or prior test examples; our study examines whether oracles can be inferred from requirements alone.

\section{Approach}
\begin{figure}[t]
    \centering
    \includegraphics[width=\linewidth]{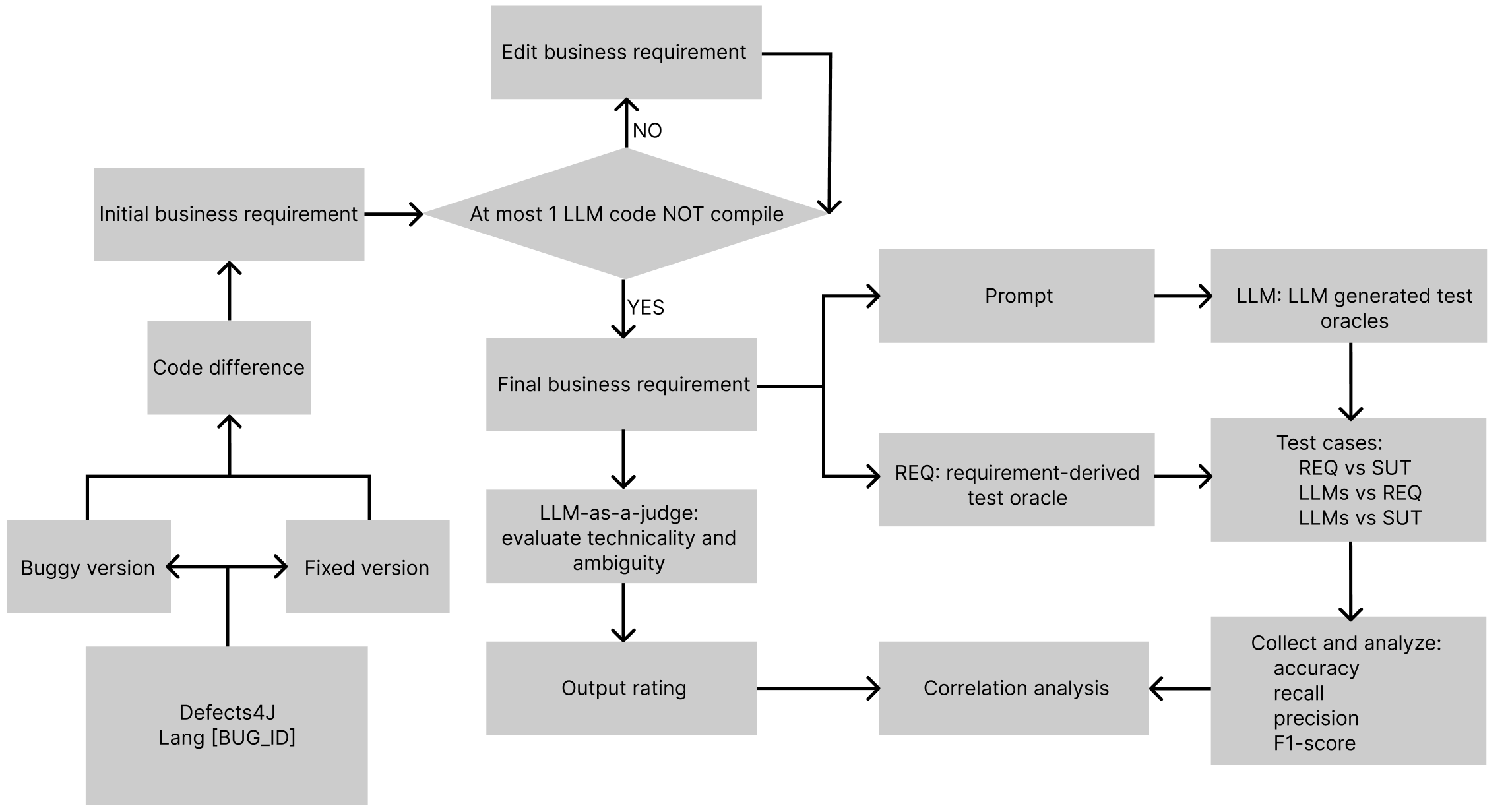}
    \caption{Workflow of the proposed approach.}
    \label{fig:screenshot}
\end{figure}
Our approach evaluates a requirements-only oracle generation setting: the LLM receives a natural-language business requirement and minimal method-level structure needed for compilation, but does not receive source code, buggy-to-fixed diffs, or example input-output pairs.\\
Our pipeline (Figure~\ref{fig:screenshot}) takes real bugs from Defects4J~\cite{just2014defects4j}, derives a natural-language business requirement from each buggy-to-fixed diff, constructs a ground-truth REQ, and prompts five LLMs to generate their own oracles from the requirement alone, without access to source code or example inputs. The REQ oracle represents our manually constructed reference interpretation of the intended requirement-level behavior implied by the fix. Generated oracles are evaluated against both REQ and the SUT using a systematically constructed test suite. Comparing against REQ measures whether the LLM captures the intended specification, whereas comparing against SUT measures whether the generated oracle matches the observed implementation behavior.

\section{Implementation}
Our implementation is based on the Defects4J benchmark~\cite{just2014defects4j}, focusing specifically on the \texttt{Lang} project and Bugs~1 and 3--11 (10 bugs in total), forming a pilot study, enabling detailed oracle construction and bug analysis. Full prompt templates for all bugs are provided in the replication package. 
\subsection{Identify Bug Behavior}
For each bug, we check out both the buggy and fixed versions using the following commands:

\begin{lstlisting}
defects4j checkout -p Lang -v [BUG_ID]b -w Lang_[BUG_ID]_buggy
defects4j checkout -p Lang -v [BUG_ID]f -w Lang_[BUG_ID]_fixed
\end{lstlisting}

We then compute the behavioral change by running a recursive unified diff over the two versions:

\begin{lstlisting}
diff -ru Lang_[BUG_ID]_buggy Lang_[BUG_ID]_fixed \
  > diff_Lang_[BUG_ID].txt
\end{lstlisting}

The generated difference file reflects how the bug was repaired and serves as the basis for understanding the intended corrected behavior.

\subsection{Business Requirement Construction}
Using \texttt{diff\_Lang\_[BUG\_ID].txt}, we manually derive a corresponding business requirement. Our objective is to express the expected system behavior in a non-technical, natural-language form without exposing code-level details. These requirements intentionally retain some ambiguity to mimic how stakeholders typically describe intended functionality.

\subsection{Standard Test Oracle (REQ)}
Given the code difference and the business requirement, we construct a requirement-derived test oracle, denoted as \textbf{REQ}. This oracle captures the correct behavior implied by the fix and serves as the reference oracle for comparison.

\subsection{Prompt Construction}
Each LLM prompt contains four structured components:
\begin{itemize}
    \item \textbf{Structure}: package name, class name, return type, and method signature.
    \item \textbf{Content Restrictions}: prohibitions against using imports or external libraries.
    \item \textbf{Output Format}: instructing the model to output Java code only.
    \item \textbf{Background Requirement}: the business requirement derived earlier.
\end{itemize}

Minimal implementation details (e.g., class names) are provided solely to eliminate the need for post-processing and ensure all outputs are evaluated exactly as generated.

\subsection{LLM Oracle Generation}
We evaluate five LLMs: DeepSeek-V3, Gemma-3n-E4B-it, Llama-3-70B-chat-hf, Mistral-7B-Instruct-v0.2, and Qwen3-Coder-480B-A35B-Instruct-FP8.  
All models are queried through the Together~AI API without any manual modification of their outputs.

\subsection{Test Case Construction and Evaluation}
For each bug, we construct a suite of test cases derived from the business requirement and the REQ oracle. Each test case is executed against three standards:

\begin{enumerate}
    \item \textbf{SUT}: the actual implementation in Defects4J.
    \item \textbf{REQ}: the requirement-derived oracle.
    \item \textbf{LLM-Oracle}: the oracle produced by each LLM.
\end{enumerate}

This enables three key comparisons:
\begin{itemize}
    \item \textbf{REQ vs.\ SUT}: Does the implementation satisfy the requirement?
    \item \textbf{LLM vs.\ REQ}: Does the LLM capture the intended specification?
    \item \textbf{LLM vs.\ SUT}: Does the LLM match real program behavior?
\end{itemize}
If an LLM-generated oracle does not compile, we exclude that bug-model pair from the evaluation metrics and report the number of evaluated pairs in the summary table.

\subsection{Result Analysis}
We analyze results along three dimensions.

First, we measure the technicality and ambiguity of each business requirement using ChatGPT~5.1 as an LLM-based evaluator, due to limited tool support for requirement ambiguity analysis. The evaluation prompt is:

\begin{lstlisting}
I'm doing research on LLMs' performance to generate generalized test oracles given business requirements. Here are business requirements provided to LLMs, rate technical degree(1-5, where 1=not technical and 5=very technical) and ambiguous degree(1-5, where 1=not ambiguous and 5=very ambiguous): 
[bug_id: business requirement]

\end{lstlisting}

Second, we compute accuracy, precision, recall, and F1-score for each bug and report macro-averaged values across all models and standards.

Finally, we compute Pearson correlation coefficients and 95\% confidence intervals to explore relationships between requirement attributes
(technical degree, ambiguity) and LLM oracle accuracy. This allows us to assess whether requirement characteristics systematically influence LLM performance.

\section{Evaluation}
We evaluate three aspects: (1) characteristics of the business requirements, (2) per-bug LLM oracle performance, and (3) correlation between requirement attributes and oracle accuracy.

\subsection{Characteristics of the Business Requirements}

We scored each requirement on \textit{technical degree} and \textit{ambiguity degree} (1--5) using ChatGPT~5.1 as an LLM-as-a-judge. Table~\ref{tab:req_characteristics} shows the results. Requirements range from highly technical with moderate ambiguity (Bugs~1, 6) to simple and unambiguous (Bugs~9, 10), providing a small but varied pilot benchmark for requirement-driven oracle generation.

\begin{table*}[t]
\centering
\caption{Technicality and Ambiguity of Business Requirements (Rated by GPT-5.1)}
\label{tab:req_characteristics}

\setlength{\tabcolsep}{3pt}
\renewcommand{\arraystretch}{1.0}
\begin{tabular}{p{0.07\textwidth} p{0.13\textwidth} p{0.13\textwidth} p{0.50\textwidth}}
\hline
\textbf{Bug} &
\makecell[l]{\textbf{Technical}\\\textbf{Degree}\\\textbf{(1--5)}} &
\makecell[l]{\textbf{Ambiguity}\\\textbf{Degree}\\\textbf{(1--5)}} &
\textbf{Rationale (Short)} \\
\hline
Bug1  & 5 & 3 & Highly technical numeric-type and boundary rules; moderate ambiguity about exact promotion order and edge-case definitions. \\ \hline
Bug3  & 4 & 4 & Conceptually clear but vague thresholds; moderately technical numeric representation logic. \\ \hline
Bug4  & 3 & 3 & Medium technicality; ambiguity about visible characters and Unicode normalization. \\ \hline
Bug5  & 4 & 3 & Moderately technical string-format parsing; ambiguity in malformed patterns. \\ \hline
Bug6  & 5 & 2 & Very technical Unicode iteration; low ambiguity due to clear rules. \\ \hline
Bug7  & 3 & 2 & Straightforward error-handling rules with low ambiguity. \\ \hline
Bug8  & 4 & 4 & Technical timezone/calendar behavior with ambiguous DST logic and missing-field handling. \\ \hline
Bug9  & 3 & 2 & Logic is simple; low ambiguity beyond definition of position handling. \\ \hline
Bug10 & 2 & 1 & Extremely clear; minimal technicality and almost no ambiguity. \\ \hline
Bug11 & 3 & 2 & Some technical nuance about bounds and defaults; low ambiguity due to explicit rules. \\
\hline
\end{tabular}
\end{table*}

\subsection{Performance of LLM-generated Oracles}
In this section, we first examine LLM performance on the ten bugs, detailing the observed behaviors relative to the underlying code changes. We then summarize the overall results by reporting the average macro-level metrics and highlighting general performance trends.

Table~\ref{tab:per_bug_summary} summarises per-bug outcomes. Bug~8 is the easiest (all models reach perfect accuracy); Bug~3 is the hardest (multi-step digit-counting logic defeats every model). Bugs involving structured lookup rules (Bug~4) or simple conditional exceptions (Bug~9) are handled well by most models, while numeric-type promotion (Bugs~1, 3) and subtle whitespace handling (Bug~10) prove consistently difficult. Mistral-7B is the weakest model across bugs, occasionally producing non-compiling output; Gemma-3n fails specifically on Unicode code-point iteration (Bug~6).

\begin{table}[t]
\centering
\caption{Per-bug performance summary (SUT target, macro-F1 range across 5 LLMs).}
\label{tab:per_bug_summary}
\setlength{\tabcolsep}{4pt}
\begin{tabular}{l l p{4.2cm}}
\hline
\textbf{Bug} & \textbf{F1 range} & \textbf{Key challenge} \\
\hline
Bug~1  & 0.40--0.92 & Hex digit-count type promotion \\ \hline
Bug~3  & 0.20--0.55 & Decimal digit-count type selection \\ \hline
Bug~4  & 0.60--1.00 & Unicode lookup table (easy for most) \\ \hline
Bug~5  & 0.82--1.00 & Locale-format validation \\ \hline
Bug~6  & 0.50--1.00 & Code-point index iteration \\ \hline
Bug~7  & 0.70--1.00 & Double-dash rejection rule \\ \hline
Bug~8  & 1.00--1.00 & Timezone calendar lookup (trivial) \\ \hline
Bug~9  & 0.75--1.00 & Region-end exception guard \\ \hline
Bug~10 & 0.40--0.85 & Subtle whitespace non-removal \\ \hline
Bug~11 & 0.80--1.00 & Boundary \texttt{end<=start} check \\
\hline
\end{tabular}
\end{table}

Table~\ref{tab:avg_llm_performance} reports macro-averaged metrics. DeepSeek-V3 is the top performer; Llama-3 and Qwen-3 form a solid second tier. Gemma-3n and Mistral-7B trail significantly. Across all models (Gemma-3n excepted), REQ-target scores exceed SUT-target scores, confirming that LLM-generated oracles align more naturally with the requirement specification used in prompting than with the underlying implementation behavior. The Count column shows how many bug cases were successfully evaluated for each model and target after excluding non-compiling oracle outputs.

\begin{table*}[t]
\centering
\caption{Average LLM Performance Across 10 Bugs}
\resizebox{\textwidth}{!}{
\begin{tabular}{l l c c c c c}
\hline
\textbf{LLM} & \textbf{Target} & \textbf{Avg Acc} & \textbf{Avg MacroP} &
\textbf{Avg MacroR} & \textbf{Avg MacroF1} & \textbf{Count} \\
\hline
REQ-based   & SUT & 0.9580 & 0.7942 & 0.7952 & 0.7929 & 10 \\ \hline
Deepseek-V3 & SUT & 0.8590 & 0.7500 & 0.7569 & 0.7476 & 10 \\
Deepseek-V3 & REQ & 0.8892 & 0.8217 & 0.8280 & 0.8179 & 10 \\ \hline
Gemma-3n    & SUT & 0.7100 & 0.6659 & 0.6544 & 0.6356 & 10 \\
Gemma-3n    & REQ & 0.7064 & 0.6104 & 0.6113 & 0.5947 & 10 \\ \hline
Llama-3     & SUT & 0.8455 & 0.7130 & 0.7259 & 0.7140 & 10 \\
Llama-3     & REQ & 0.8695 & 0.7986 & 0.7844 & 0.7797 & 10 \\ \hline
Mistral-7B  & SUT & 0.6440 & 0.5699 & 0.5825 & 0.5428 & 9 \\
Mistral-7B  & REQ & 0.6617 & 0.5930 & 0.5831 & 0.5600 & 9 \\ \hline
Qwen-3      & SUT & 0.8077 & 0.6992 & 0.7188 & 0.7030 & 10 \\
Qwen-3      & REQ & 0.8379 & 0.7725 & 0.7983 & 0.7779 & 10 \\
\hline
\end{tabular}
}
\label{tab:avg_llm_performance}
\end{table*}

\subsection{Correlation Between Requirement Attributes and Oracle Accuracy}

\begin{figure*}[t]
    \centering
    \includegraphics[width=\textwidth]{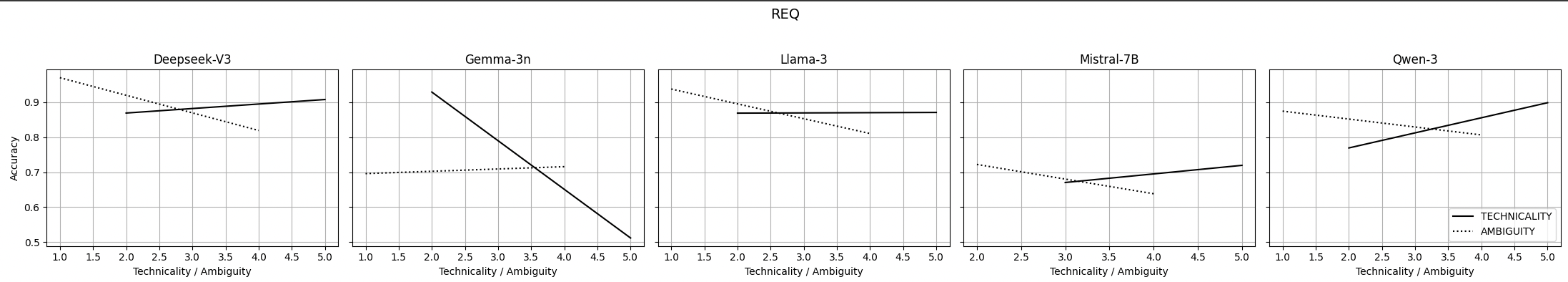}
    \includegraphics[width=\textwidth]{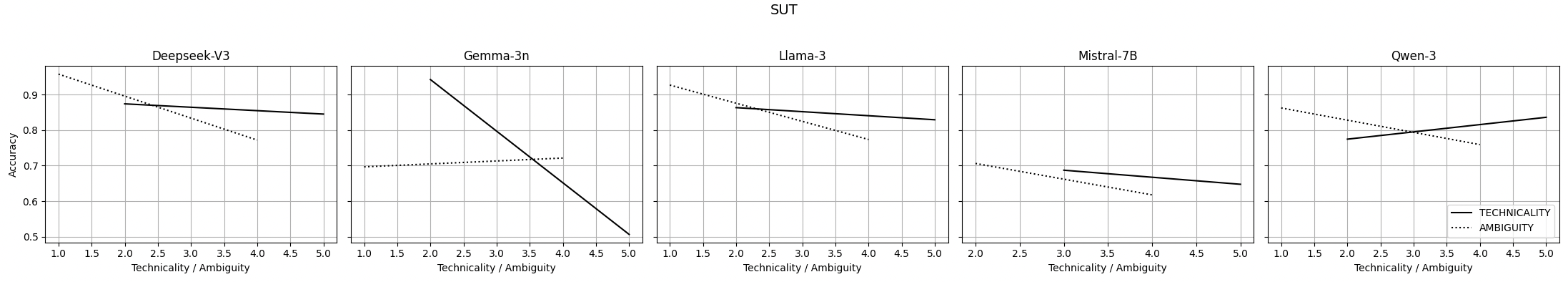}
    \caption{Pearson correlation between accuracy and requirement attributes under REQ (top) and SUT (bottom) targets across all five LLMs.}
    \label{fig:sut_req_acc}
\end{figure*}

For each of the ten bugs, we compute the Pearson correlation coefficient ($r$) and 95\% CI between each requirement attribute (technicality; ambiguity) and per-model oracle accuracy, under both SUT and REQ targets. Figure~\ref{fig:sut_req_acc} visualizes the results.


All $r$ values are near zero (technicality: $-0.50$ to $+0.24$; ambiguity: $-0.36$ to $+0.03$) and all 95\% CIs include zero under both evaluation settings. In this small dataset, no model exhibits a consistent directional trend. We conclude that, within this dataset, \textbf{neither technicality nor ambiguity reliably predicts LLM oracle accuracy in this pilot dataset}. Other factors, such as pretraining coverage of the relevant algorithmic domain or semantic specificity of the required behavior, appear to dominate oracle correctness.

\section{Discussion}
This section interprets the findings of our three research questions, discusses threats to validity, and outlines implications for researchers and practitioners.

\subsection{RQ Answers}

\textit{\textbf{RQ1: Generalization of Requirement-Driven LLM Test Oracles.}}
LLMs can generate oracles that generalize to unseen test cases to a limited but measurable extent; performance varies substantially across bugs, with numerical-type promotion and digit-counting logic posing the greatest challenge, while structured rule-based bugs are handled reliably.

\textbf{\textit{RQ2: Comparison Against REQ-Derived and SUT Oracles.}}
LLM-generated oracles score consistently higher against the REQ oracle than against the SUT, confirming that requirement-prompted LLMs model the intended specification more faithfully than the underlying implementation. This gap suggests that requirement-prompted LLMs behave more like specification-level oracle generators than implementation-level ones. DeepSeek-V3 leads across both targets; Mistral-7B trails, with occasional non-compiling output.

\textbf{\textit{RQ3: Influence of Requirement Technicality and Ambiguity.}}
Pearson correlations between requirement technicality or ambiguity and LLM oracle accuracy are near zero across all models and targets (all 95\% CIs include zero), indicating that surface-level requirement complexity does not reliably predict oracle correctness within this dataset.

\subsection{Interpretation of Findings}

\textbf{\textit{Partial generalization is achievable but inconsistent.}}
Our results show that modern LLMs can infer correct oracle logic from natural-language business requirements and apply it to unseen test inputs, but only partially. The wide spread in per-bug accuracy, from near-perfect on structured, rule-based bugs (e.g., Bug~8, Bug~11) to consistently poor performance on digit-counting or numeric-type promotion bugs (e.g., Bug~3), suggests that oracle quality depends more on the semantic nature of the required behavior than on any surface-level property of the requirement text. Bugs whose correct behavior maps closely to common programming idioms present in LLM training data are handled well; bugs requiring multi-step numeric reasoning or deep domain knowledge of type promotion rules are not.

\textbf{\textit{Requirement-level prompting creates a REQ--SUT gap.}}
A consistent finding across all five models is that LLM-generated oracles score higher against the REQ than against the SUT. This is an expected but important observation: because LLMs are prompted with the business requirement as the sole source of truth, they learn the specification's language, not the implementation's behavior. This means that LLM-generated oracles are more useful for \emph{requirement conformance checking} than for regression testing of existing systems. Practitioners should treat these two use cases as distinct when selecting oracle generation strategies.

\textbf{\textit{Requirement complexity does not predict model accuracy in this dataset.}}
The absence of a detectable linear correlation between technicality or ambiguity ratings and oracle accuracy has a nuanced implication. It provides little support for the simple hypothesis that ``harder requirements produce worse oracles,'' but it does not mean that requirement properties are irrelevant. Rather, it suggests that the relationship, if any, is nonlinear or moderated by factors such as the semantic distance between the requirement text and the LLM's training distribution. Future work should explore richer characterizations of requirement difficulty, such as the number of logical conditions or the presence of numeric constraints, rather than coarse ordinal ratings.

\subsection{Threats to Validity}

\textbf{\textit{Internal.}}
Business requirements and REQ oracles were derived manually from buggy-to-fixed diffs, introducing author subjectivity; different phrasing could shift LLM outputs. Technicality/ambiguity ratings from a single LLM judge (ChatGPT~5.1) carry no inter-rater reliability estimate. Test case suites are also manually constructed; a different suite could shift per-bug metrics. Our handling of non-compiling LLM outputs may also affect aggregate metrics, since excluding such outputs differs from treating them as oracle failures.

\textbf{\textit{External.}}
The study covers 10 bugs from one Defects4J component (\texttt{Lang}) in Java, concentrated in string and numeric utilities. Generalisation to other languages, bug types, or requirement styles is untested. LLM results may shift with model updates.

\textbf{\textit{Construct.}}
The REQ oracle is a single human interpretation and may not be unique. Accuracy against REQ/SUT does not fully capture practical oracle utility (e.g., false-positive rates in CI).

\subsection{Implications}

\textbf{\textit{For researchers.}}
This study provides an early, reproducible benchmark for requirement-driven oracle generation, a problem that existing literature addresses far less than input generation or assertion synthesis from source code~\cite{molina2025test,barr2015oracle}. The pipeline, business requirements, and evaluation test cases are made available to support follow-on studies with larger bug corpora or finer-grained requirement representations.

\textbf{\textit{For practitioners.}}
LLMs such as DeepSeek-V3 and Llama-3 already achieve macro-averaged accuracy above 0.84 against REQ, suggesting they are viable assistants for requirement-driven oracle drafting. However, developers should not rely on LLM-generated oracles as a substitute for manually reviewed assertions in safety-critical or numerically intensive code paths, where model accuracy drops significantly.

\section{Conclusion}
This study piloted whether LLMs can generate requirement-driven test oracles that generalize from business requirements alone, using real bugs from Defects4J across ten bugs, five LLMs, and two oracle targets.

Our findings show that LLMs are capable of producing test oracles that capture substantial portions of intended behavior. Models such as Deepseek-V3, Llama-3, and Qwen-3 achieve high accuracy and reasonable F1 scores on unseen inputs, indicating that the problem is \emph{partially solved}, particularly when required behavior is structured and well-defined. However, performance varies significantly: models struggle with nuanced numerical rules, Unicode handling, and edge-case validations, and weaker models like Gemma-3n and Mistral-7B often fail to produce compilable or correct outputs. LLM-generated oracles align more closely with the requirement-derived (REQ) oracle than with SUT behavior, confirming that LLMs more readily model \emph{intended} specification than exact implementation logic.

Our correlation analysis finds no meaningful relationship between requirement attributes (technicality and ambiguity) and LLM accuracy, suggesting that limitations arise from semantic or logical task difficulty rather than linguistic requirement complexity. \textbf{In summary}, LLMs provide a promising but incomplete solution to requirement-driven oracle generation, useful as assistive tools in automated testing, but not yet dependable enough to replace manually engineered oracles where high precision and semantic correctness are required.

\section*{Data Availability}
The replication package for this study, including the business requirements derived for each bug, the REQ oracles, the LLM-generated oracles, the test case suites, and the evaluation scripts, is available at: \url{https://figshare.com/s/50c240d183532de44c7a}.
The Defects4J benchmark used to derive bugs is publicly available at \url{https://github.com/rjust/defects4j}~\cite{just2014defects4j}.
LLM outputs were obtained via the Together~AI API; model names and versions are documented in the replication package.

\bibliography{sigproc}

\end{document}